\documentstyle[twoside, epsf]{article}

\input ibvs2.sty 

\begin{document}
\IBVShead{xxxx}{03 Nov 2014}

\IBVStitletl{New variables in M5 (NGC 5904) and some identification corrections}

\begin{center}
\IBVSauth{Arellano Ferro, A.$^1$; Bramich, D.M.$^2$, Giridhar, S.$^3$, Luna,
A.$^{1}$, Muneer, S.$^3$}
\end{center}

\IBVSinst{Instituto de Astronom\'ia, Universidad Nacional Aut\'onoma de M\'exico,
Ciudad Universitaria CP 04510, M\'exico: armando@astro.unam.mx; aluna@astro.unam.mx}
\IBVSinst{Qatar Environment and Energy Research Institute, Qatar Foundation, P.O. Box
5825, Doha, Qatar: dan.bramich@hotmail.co.uk}
\IBVSinst{Indian Institute of Astrophysics, Koramangala 560034, Bangalore, India:
giridhar@iiap.res.in}

\begintext

\section{Introduction}

The bright globular cluster M5 (NGC 5904) has been the subject of many variable star
searches for more than a hundred years. The first variables were discovered by Bailey
(1902). The catalogue of variable stars in globular clusters (CVSGC; Clement et al.
2001) lists 169 variables, mostly of the RR Lyrae type, with 5 SX Phe stars, one
W Virginis star (CW), one RV Tau, one (possibly two) eclipsing binaries, and one U
Gem type star. However, there
are also a
number
of uncertain classifications and some variables have an unknown type, or it is not
even clear if they are truly variable. A new study of the variable stars in M5 is
therefore pertinent.

As part of our program of CCD time-series observations of variable star populations in
globular clusters (GC), we performed CCD $V$ and
$I$ photometry of the globular cluster M5. Difference
image analysis (DIA) has proven to be very
efficient in identifying variable stars even in the crowded central regions of
GCs (e.g. Arellano Ferro et al. 2013 and references therein).
Exploration of our collection of light curves of all stars in the field of our images
down to $V \sim$ 18.5 mag allowed us to identify twelve variables not previously
detected; one SX Phe and eleven semi-regular variables (SR). In the present
note, we report on their
identifications, equatorial coordinates, ephemerides, and light curves. 
We argue that the known variable V155, previously classified as RRc, is in fact a
contact eclipsing binary or EW.
Furthermore, we have explored the light curves of a group of stars whose variability
has not
been confirmed and that are marked as probable non-variables in the CVSGC. Finally,
we offer detailed identifications for some of the known variables in crowded regions
that
were misidentified in previous studies. We shall also address the cases of the
cataclysmic variable or U Gem type V101 and of the variable blue straggler V159.

\section{Observations and reductions}
The observations were acquired on 11 nights between 29th February 2012 and 9th April
2014 with the 2.0m telescope of the Hanle
Observatory, India. A total of 385 and 384 images in $V$ and $I$, respectively, were
obtained.
Image data were calibrated using bias and flat-field correction procedures. We used
DIA to extract high-precision time-series photometry employing the {\tt DanDIA}
pipeline for the data reduction process (Bramich et al.
2013), which includes an algorithm that models the
convolution kernel matching the PSF of a pair of images of the same field as a
discrete pixel array (Bramich 2008). We have also applied a post-calibration
method developed by Bramich \& Freudling (2012) which determines appropriate
per-image magnitude
offsets to correct for errors in the fitted value of the photometric scale factor $p$.
We derived offsets of the order of $\sim0.02$ and $\sim0.03$ mag in $V$ and $I$,
respectively. The instrumental magnitudes are calculated via the difference
flux, the reference flux and the photometric scale factor by the equation;

\begin{equation}
m_{\mbox{\scriptsize ins}}(t) = 25.0 - 2.5 \log \left[f_{\mbox{\scriptsize ref}} +
\frac{f_{\mbox{\scriptsize diff}}(t)}{p(t)}
\right].
\label{eqn:mag}
\end{equation}

The difference fluxes $f_{\mbox{\scriptsize diff}}$ are measured by
scaling the known PSF to the
difference images at the position of each star. Since the constant stars have been
fully subtracted in the difference images, the difference fluxes for the variables are
very precise. The reference fluxes $f_{\mbox{\scriptsize ref}}$ are, however,
measured on the reference image
by PSF fitting and they have the potential to suffer from the usual problems caused by
blending. For the variables in the most crowded parts of the reference image, where
the probability of blending is high, the brightness of a variable star may be
overestimated, and its amplitude underestimated (see Section 2.3 of Bramich et al.
2011 for a more in-depth discussion of the caveats of DIA).

The instrumental magnitudes were transformed to the standard Johnson-Kron-Cousins
magnitudes using secondary photometric standards in the field of view (FoV)
from Stetson (2000) covering the full range of colours. 

All of our $VI$ photometry for the stars discussed in this paper is provided
in Table \ref{tab:vi_phot}. Just a small portion of this
table is given in the printed version of this paper, while the full table is
only available in electronic form.

\begin{table*}
\footnotesize{
\caption{Time-series $V$ and $I$ photometry for all stars discussed in this paper. The
standard $M_{\mbox{\scriptsize std}}$ and
instrumental $m_{\mbox{\scriptsize ins}}$ magnitudes are listed in columns 4 and 5,
respectively, corresponding to the variable star in column 1. Filter and epoch of
mid-exposure are listed in columns 2 and 3, respectively. The uncertainty on
$m_{\mbox{\scriptsize ins}}$ is listed in column 6, which also corresponds to the
uncertainty on $M_{\mbox{\scriptsize std}}$. For completeness, we also list the
quantities
$f_{\mbox{\scriptsize ref}}$, $f_{\mbox{\scriptsize diff}}$ and $p$ from
Eq.~\ref{eqn:mag}
in columns 7, 9 and 11, along with the uncertainties $\sigma_{\mbox{\scriptsize
ref}}$ and $\sigma_{\mbox{\scriptsize diff}}$ in columns 8 and 10. This is an extract
from the full table, which is available with the electronic version of the article
(see Supporting Information).}
\centering
\begin{tabular}{ccccccccccc}
\hline
Variable &Filter & HJD & $M_{\mbox{\scriptsize std}}$ &
$m_{\mbox{\scriptsize ins}}$
& $\sigma_{m}$ & $f_{\mbox{\scriptsize ref}}$ & $\sigma_{\mbox{\scriptsize ref}}$ &
$f_{\mbox{\scriptsize diff}}$ &
$\sigma_{\mbox{\scriptsize diff}}$ & $p$ \\
Star ID  &        & (d) & (mag)                        & (mag)                       
& (mag)        & (ADU s$^{-1}$)               & (ADU s$^{-1}$)                    &
(ADU s$^{-1}$)                &
(ADU s$^{-1}$)                     &     \\
\hline
V23& V & 2455987.37467 & 14.402 & 15.567 & 0.002 &   5984.932&  11.137& -51.549& 
 9.416& 1.0195\\
V23& V&  2455987.37903 & 14.404 &15.569  &0.002  &  5984.932&  11.137&   -59.403&   
9.521& 0.9697\\
\vdots   & \vdots & \vdots  & \vdots & \vdots & \vdots & \vdots   & \vdots & \vdots &
\vdots & \vdots \\
V23& I&  2455987.37248 & 13.353 & 14.622&  0.002&   14205.530&  25.101& -44.978
& 27.407 &1.0588\\
V23& I&  2455987.37686&  13.370 & 14.639&  0.002 &  14205.530&  25.101& -278.790  &
30.268 &1.0423\\
\vdots   & \vdots & \vdots  & \vdots & \vdots & \vdots & \vdots   & \vdots & \vdots  &
\vdots & \vdots \\
V25& V & 2455987.37467&  14.632&  15.826&  0.003& 2449.232&  15.883&   +2267.438&  
12.352& 1.0195\\
V25& V & 2455987.37903&  14.625&  15.819&  0.003& 2449.232&  15.883&   +2186.565&  
11.576& 0.9697\\
\vdots   & \vdots & \vdots  & \vdots & \vdots & \vdots & \vdots   & \vdots & \vdots  &
\vdots & \vdots \\
V25& I&  2455987.37248 & 14.253&  15.542&  0.004&    3901.695&  29.242&   +2298.204&  
26.254& 1.0588\\
V25& I&  2455987.37686 & 14.310&  15.599&  0.005&    3901.695&  29.242&   +1939.003&  
29.961& 1.0423\\
\vdots   & \vdots & \vdots  & \vdots & \vdots & \vdots & \vdots   & \vdots & \vdots  &
\vdots & \vdots \\
\hline
\end{tabular}
\label{tab:vi_phot}
}
\end{table*}

\section{Exploration of suspected non-variables in the CVSGC}

In the CVSGC there are 23 stars
classified as (probably) non-variables or "constant" (CST, CST? or ?); these are
V22, V23, V46, V48,
V49, V51, V124, V136, V138, V140, V141, V143-V154. Except for V22 and V141, which are
outside of the FoV of our
images, we have $VI$ light curves for all of them. We have carried out a
quick
exploration of the light curves to comment on their possible variability or otherwise.

\IBVSfig{10.0cm}{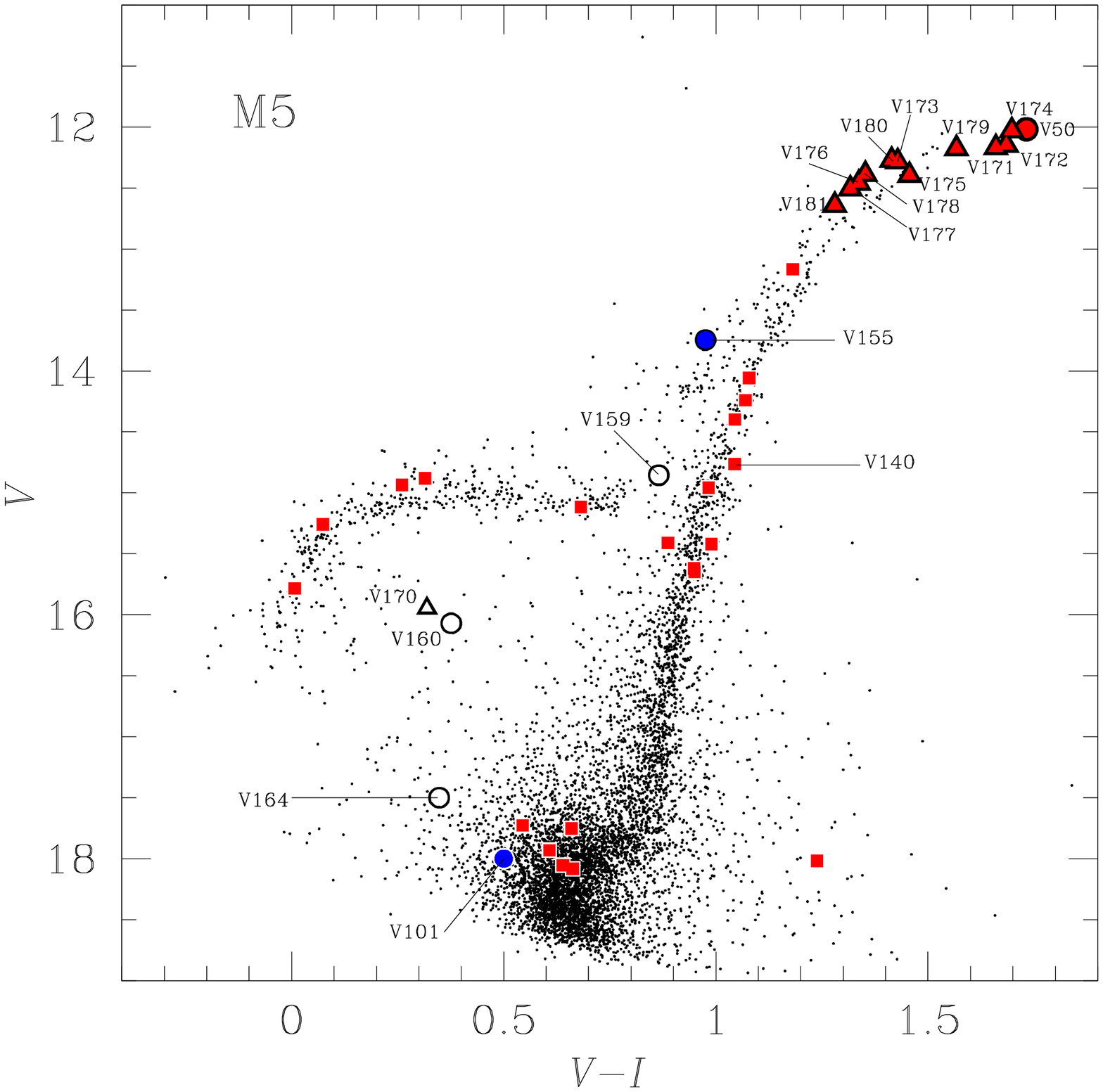}{Colour-magnitude
diagram of M5 with the new variables marked with triangles. The colour code is: 
empty symbols for blue straggler variables, the position of binary V159 is biased
since it is
heavily blended; red triangles for tip of the RGB variables; blue filled circles for
two previously known variables V101 and V155. The cataclysmic variable V101 is
plotted at its
approximate position during outburst. No other known variables are shown.  Red
squares are stars listed as variables but whose variability has not been
confirmed. See the text for a detailed discussion.}

\IBVSfig{10.0cm}{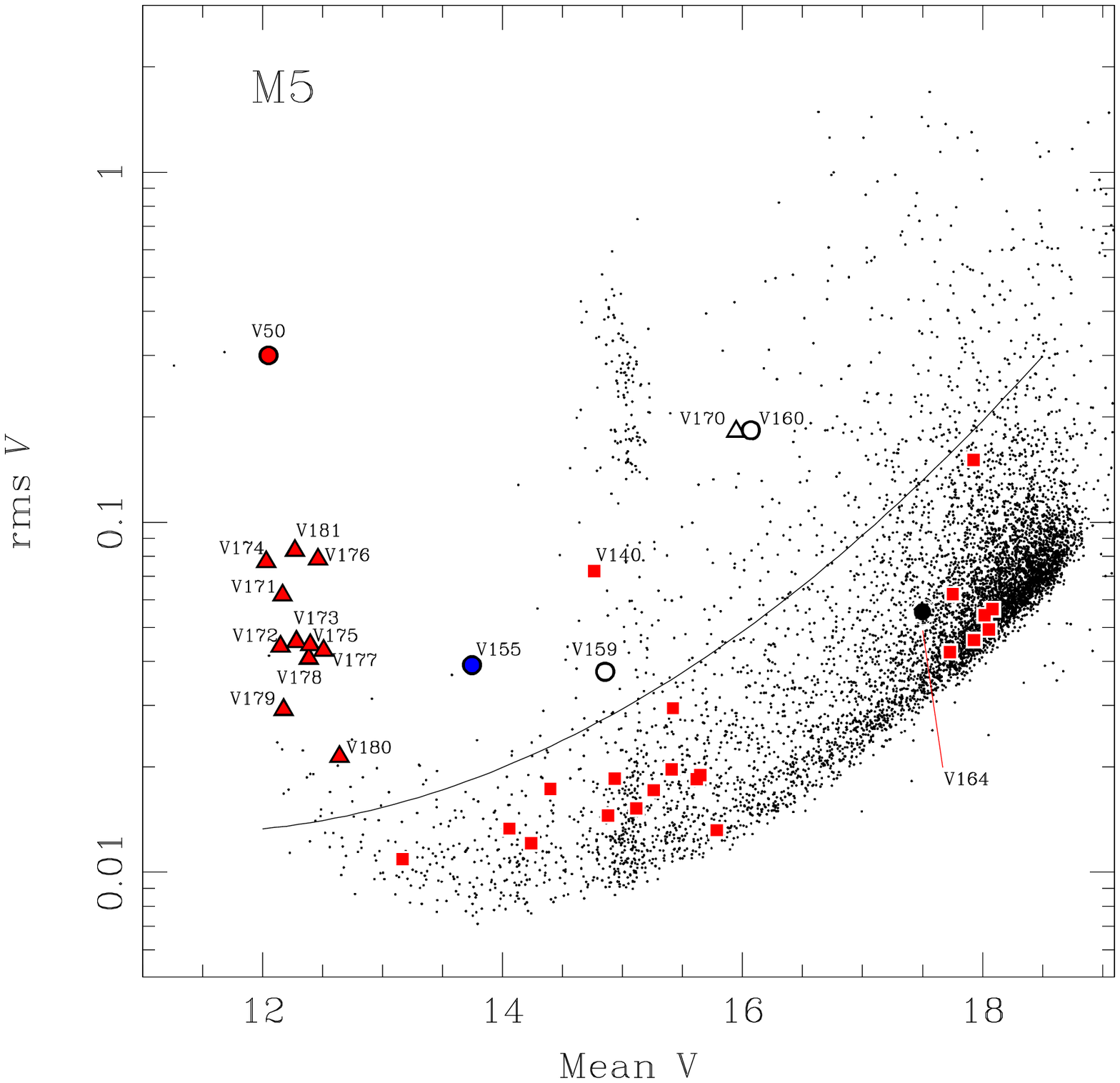}{The rms $V$ magnitude deviations calculated
for 8431 stars in our FoV of M5 as a function of mean magnitude $V$. The sybols and
colour code are as in Fig 1. The continuous line represents an arbitrary threshold for
variability detectability (see text in $\S$ 3).}

\begin{table}[!ht]
\caption {Mean magnitudes and rms for stars whose variability is
not confirmed. Local rms refers to the upper limit of the main rms distribution for a
given value of
$<V>$, represented by the continuous line in Fig. 2.}
\begin{center}
\begin{tabular}{@{}cccc|cccc@{}}
\hline
Var ID   & $<V>$&rms $V$ &local rms& Var ID   & $<V>$&rms $V$ &local rms\\
\hline
V23  &14.40&0.017&0.023 &V145&15.26&0.017&0.033\\
V46 &17.93&0.150&0.184 &V146 &15.65&0.019&0.040\\
V48 &14.24&0.012&0.022 &V147 &15.12&0.015 &0.031 \\
V49 &15.78&0.013&0.043 &V148 &15.62&0.018 &0.040 \\
V51 &14.06&0.013&0.020&V149 &17.75&0.062&0.160\\
V124&14.88&0.015&0.028 &V150&18.08&0.056 &0.208 \\
V136 &14.93 &0.018&0.028&V151&18.02&0.054&0.198\\\
V138 &13.17&0.011&0.016 &V152&17.72&0.043&0.160\\
V140 &14.76&0.072&0.028 &V153&18.05&0.049&0.203\\
V143 &15.42&0.029&0.036 &V154&17.93&0.046& 0.184\\
V144&15.41&0.020&0.036  & & & & \\
\hline
\end{tabular}
\end{center}
\end{table}

Firstly we have identified the 21 stars in our FoV on the colour-magnitude diagram
(CMD) of Fig. 1 and the RMS diagram of Fig. 2. All of them  
are marked with  red squares.  In the RMS diagram we draw an arbitrarily set
line above which all variables seem to fall and hence it can serve as a guide of
detectability when
judging the variabilty of a given candidate. While true variables are expected
to have significantly larger rms values than this upper limit, we
note however that non-true variables may lie above that limit if they are near a true
variable due to flux contamination (e.g. V140, see below), or that true variables may
be found below that limit, particularly those of very small amplitude
(e.g. the SX Phe star V164). Thus, individual explorations of the light curves of
specific cases is required. In Table 2 we list the mean magnitudes
and rms values in the $V$ light curves. In the last column we list the value of the
upper limit rms corresponding to the mean magntiude.
Except for V140, all the stars listed above fall below the threshold.
From the individual explorations we found that V140 does in fact show some
variations.
However we argue that these are the due to flux contamination by a nearby variable
(V175) as it will be discussed in $\S$ \ref{newvar}. For the rest, we found no
signs of variability confirming their classification  as
non-variables (or CST) in the CVSGC.

\section{Comments, identification and classification correction, for some
previously known variables}

During the process of variable star identification we noticed that stars V25, V36,
V53,
V74, V102 and V108 are all very close to a neighbour of similar brightness or much
brighter. While
checking the finding charts of the discovery papers, we found that their
identifications are dubious or definitely wrong, mainly due to the fact that the stars
are not resolved in
old plates and/or that they are close to the cluster central regions. Here we offer a
precise identification and a few comments on each variable. 
We have confirmed the variable nature of these stars by phasing the light curves
of the two candidates and by blinking the difference images. It was also
noted that the equatorial coordinates in the CVSGC of the variable V140 point to a
non-variable star. We address the cases of V50 whose variability has not been
clearly established and of V155 that needs a reclassification. To avoid confusion 
in future work we include here in Table 3 the correct equatorial coordinates of all
these stars. Below we offer a brief comment on individual stars including the U Gem
type star V101, and binary V159.

V25 is a very close pair that in our images is heavily blended. In the
finding chart of the discovery paper (Bailey 1902), the star looks like a single one.
Careful blinking of the difference images makes it clear that the true variable is the
western star of the pair, as identified in Fig. 4.

V36 is also called V135 (see CVSGC for M5, 2014 update). The star is incorrectly
identified in the chart of Caputo et al. (1999), labelled as V135, as the
south-western star of the pair. The RRab variable is actually the north-eastern and
brighter star of the pair.

V50 sits on the tip of the RGB.  Bailey (1917) suggested a period of 106d that was not
confirmed by Oosterhoff (1941) who described the variation as irregular. Our data
suggest a
period of 107.6d, in good agreement with Bailey's result. Therefore we classify
the star as  a semi-regular late-type variable (SRA). Its light curve, phased
with the above period, is shown in Fig. 3.

V53 is not resolved in the finding chart of Bailey (1902) and not identified
afterwards.
The correct variable is the eastern star of the pair.

V74 is not resolved in the finding chart of Bailey (1902) and not identified
afterwards.
The correct variable is the western star of the pair.

V101 is a cataclysmic variable of the U Gem type. It was discovered by
Oosterhoff (1941) who classified it as SS Cyg (or dwarf nova). Two outbursts
of amplitude 2.7 mag 
within 100 days in the $V$ light curve were detected by Kaluzny et al. (1999) who
argue in favour of a short duty cycle with a characteristic time of about 3.4 hours. 
Our $VI$ light curves, displayed in the mosaic of Fig. 3, span 770 days and two
outbursts are clearly seen at HJD 2456029.4 and 2456312.5 d reaching 18.5 and 18.0
mag in $V$, and 18.0 and 17.0 mag in $I$ respectively.

V102 The identification chart in Oosterhoff (1941) shows a strong blend close to the
saturated central region that prevents an accurate identification. The authentic
variable is the SE star of the pair.

V108 The variability of this star was announced by Kadla et al. (1987) (see also
Gerachencko 1987). It was
identified by Drissen \& Shara (1998) in their {\it Hubble Space Telescope (HST)}
image but mistakenly labelled as V22.
The identification of
the star by Caputo et al. (1999), now labelled as V108, points to the wrong star to
the east of the real variable. We confirm that the
variable star is the western star of the pair, in agreement with Drissen \& Shara's
(1998) identification.

\begin{table}[!ht]
\caption {Known variables with corrected classifications,
equatorial coordinates and identifications in Fig. 4. }
\begin{center}
\begin{tabular}{@{}cccc@{}}
\hline
Var ID   & Variable type &RA(J2000.0)& Dec.(J2000.0) \\
\hline
V25 &RRab &15 18 30.98 & +02 02 42.5 \\
V36 &RRab &15 18 32.66 & +02 03 58.9 \\
V50 &SRA &15 18 36.04 &+02 06 37.8\\
V53 &RRc &15 18 37.92 & +02 05 06.8 \\
V74 &RRab &15 18 47.19 & +02 07 25.7  \\
V101&U Gem &15 18 14.51& +02 05 35.7\\
V102&RRab &15 18 34.37 & +02 04 34.4  \\
V108&RRc &15 18 33.79&+02 04 47.0 \\
V140&CST&15 18 36.18&+02 05 13.2\\
V155& EW  &15 18 33.40&+02 05 12.2\\
V159& E  &15 18 32.88&+02 04 36.5\\
\hline
\end{tabular}
\end{center}
\end{table}

V140 is identified by Caputo et al. (1999) but the equatorial coordinates given in
the CVSGC have a typo error producing some confusion in the identification and
in the variable nature of the star which is classified as a probable
non-variable. The star 
pointed to by the CVSGC at 15:18:36.18; +02:03:13.1 is not variable. The correct 
coordinates of the star identified by Caputo et al. (1999) are given in Table 3. 
However, this V140 is very close to a much brighter star and a careful blinking 
of the difference images clearly reveals that the authentic
variable is the brighter star, which we have identified as the new SRA variable V175
(see $\S 
\ref{newvar}$). Both the 
light curves for V140 and V175 are shown in the mosaic of Fig. 3. The light curve of
V140 has been contaminated by the real variations in the much brighter star V175. We
conclude that V140 is not a variable star while V175 is an SRA variable.

V155 was discovered by Drissen \& Shara (1998) and they classified it as RRc.
However, the star lies near the RGB on the CMD (see Fig. 1) and the light curve in
Fig.
3 is similar to that of an
eclipsing binary of the EW type, or contact binary, when phased with the ephemerides 
P= 0.664865 days and the epoch 245 6504.2067 d. Note that two different
depths of the minima, particularly visible in the $V$ light curve, are implied.

V159 is classified in the CVSGC as a probable eclipsing binary. The star
was identified as variable by Drissen \& Shara (1998) on their HST images and
labelled as V28. The V159 name was given by Caputo et al. (1999) on their finding
chart. We find that this star is highly blended in our images, which affects the
star's position on the CMD. We detected two clear eclipses in the $V$ band at
2455989.52 and 2456750.44 d, of about 0.15 mag depth (see Fig. 3). However, we are not
able to determine the periodicity although we confirm the star as an eclipsing binary.

\section{New variable stars in M5}
\label{newvar}

To search for new variables in the field of M5 we have used several approaches.
We isolated all stars in regions of the CMD where most variable stars in a GC tend to
be found. This includes the horizontal branch, the blue stragglers region and the
RGB. We identified all
previously known variables in the field of our images and studied in detail the light
curves of the rest of the stars. This procedure allowed us to identify a new  large
amplitude ($A_V \sim 0.6$ mag) SX Phe star V170, for which we identify only one
period. 
Then we explored the difference images for clear variations; this
approach allowed us to discover the variability of six SRA (V171, V172, and
V174-V177). 
A third approach was via the rms diagram of Fig. 2. It can be seen from this diagram 
that our photometry achieves uncertainties between 7 and 20 mmag at the bright
end. High values of rms are generally produced by variable stars. For instance the
group of stars with rms above 0.1 mag and with $V \sim$ 15 are all RR Lyrae stars
which are not
discussed in the present paper. Through this procedure we identified five
additional SRA stars V173 and V178-V181.
For some of these new SRA stars we have been able to estimate a periodicity.

The new variables,
their equatorial coordinates, periods and epochs are listed in Table~4 and their
position on the CMD and the rms diagram are displayed in Figs. 1 and 2 respectively.
Their light curves in the $V$ and $I$ bands, phased whenever possible or as a function
of HJD otherwise, are displayed in Fig. 3.

In the identification chart of Fig 4, we have marked a detailed identification for
all variable stars discussed in this paper. No effort has been made to identify the
numerous
known variables listed in the CVSGC since that will be the subject of a future
paper.

\IBVSfig{19cm}{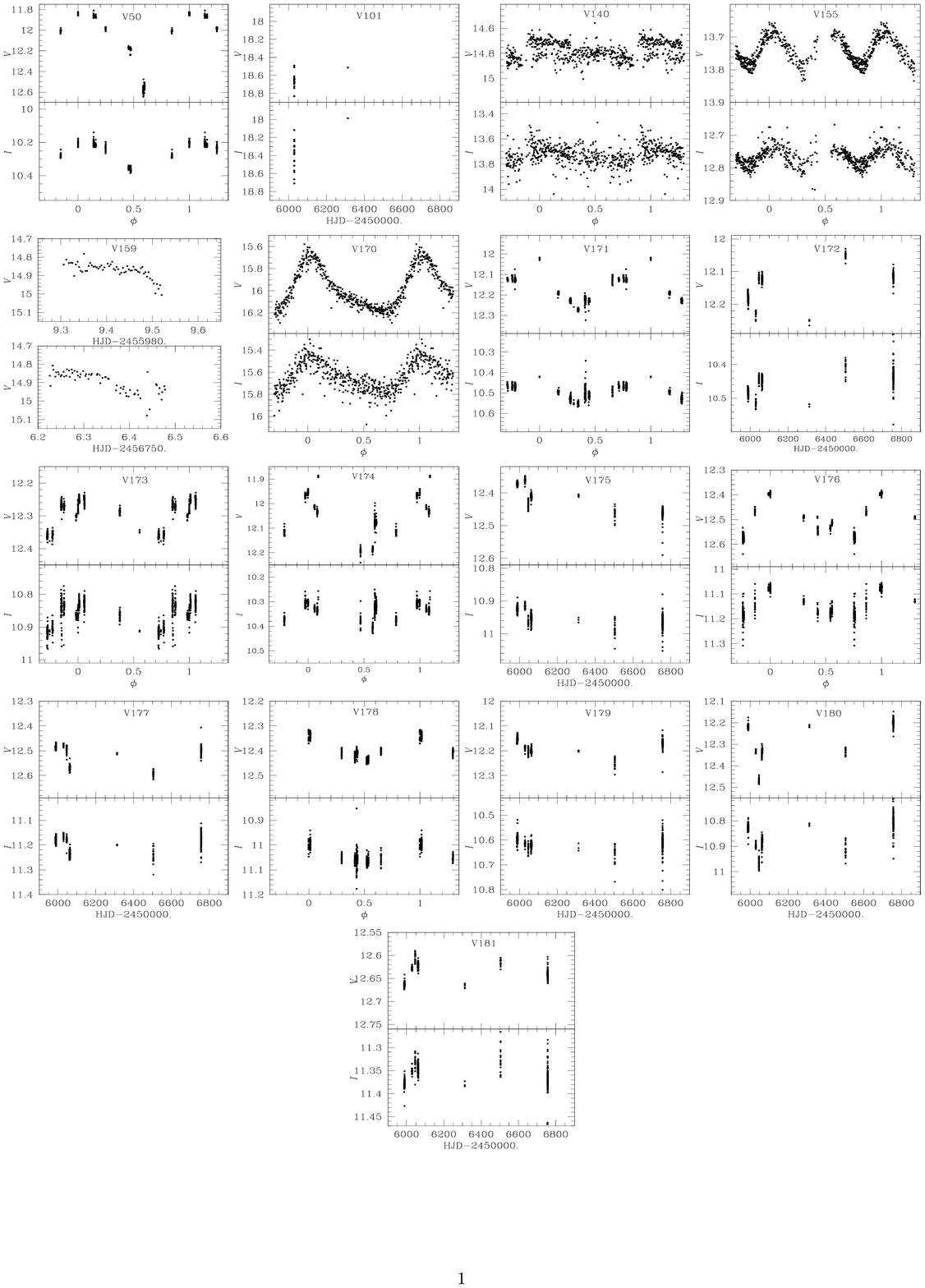}{$V$ and $I$ light curves of the variable stars
found in this work (V170-V181). V50, V101, V140, V155 and V159 are discussed in the
text. For V101 only the data during outburst are displayed. For V159 the two panels
are $V$ light curves during eclipses.
When the period is known, light curves are phased with the ephemerides in Table 4.}

\IBVSfig{18cm}{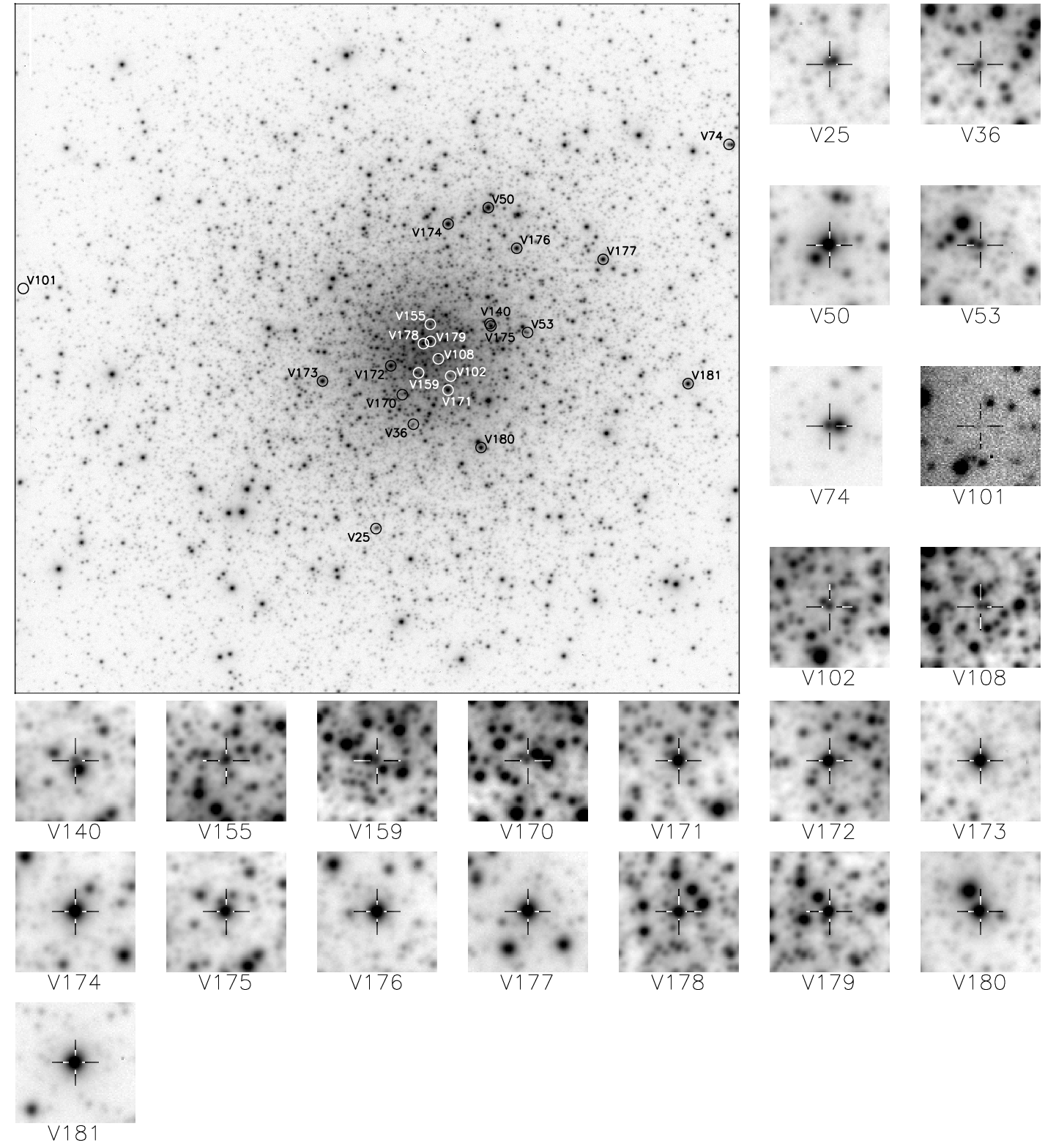}{Finding charts from our $V$ reference image;
North is up and east is to the right. The cluster image is 8.39 $\times$ 8.39
arcmin$^2$ and the individual stamps are 24.0 $\times$ 24.0 arcsec$^2$.}

\begin{table}[!ht]
\centerline{{\bf Table\,4.} New variable stars found in M5, V170-V181.}
\begin{center}
\begin{tabular}{@{}cccccccccc@{}}
\hline
Name  & Variable &RA & Dec.& Period& Epoch& $<V>$ &$<I>$ &$A_V$&
$A_I$\\
 & type &(J2000.0) &(J2000.0) & days& (-245 0000.)& mag &mag&mag&mag\\
\hline
V170&SX Phe &15 18 32.14 & +02 04 20.4 &0.089467 & 6063.3361&15.95&15.63&0.57
&0.41\\
V171& SRA  &15 18 34.26&+02 04 24.2&28.8&6312.5083&12.17&10.50&0.25&0.14\\
V172& SRA  &15 18 31.59&+02 04 41.4&--& --&12.15&10.47&0.23&0.13\\
V173& SRA   &15 18 28.42&+02 04 29.8&43.1& 6504.1686&12.28&10.86&0.13&0.13 \\
V174& SRA &15 18 34.18&+02 06 25.5&80.6&6063.4183&12.03&10.33&0.33&0.15\\
V175& SRA &15 18 36.22&+02 05 11.3& --&--&12.40&10.94&0.18&0.13\\
V176& SRA &15 18 37.38 & +02 06 08.2 &133.3&5989.3064&12.46&11.13&0.22&0.20\\
V177& SRA  &15 18 41.40&+02 06 00.9&--&--&12.51&11.19&0.13&0.10\\
V178& SRA  &15 18 33.10&+02 04 58.0&141.6&5987.4759&12.39&11.03&0.12&0.10\\
V179& SRA  &15 18 33.42&+02 04 59.6&--&--&12.18&10.61&0.12&0.11\\
V180& SRA  &15 18 35.82&+02 03 42.4&--&--&12.27&10.86&0.24&0.24\\
V181& SRA  &15 18 45.40&+02 04 30.9&--&--&12.64&11.36&0.07&0.08\\
\hline
\hline
\end{tabular}
\end{center}
\end{table}

\bigskip

ACKNOWLEDGMENTS: We are thankful to the referee Dr. \'Ad\'am S\'odor for his valuable
comments and suggestions. We acknowledge the support from DGAPA-UNAM, Mexico grant
through
projects IN104612 and IN106615 and from NPRP grant \# X-019-1-006 from the Qatar
National Research
Fund (a member of Qatar Foundation).

\bigskip
\bigskip
\bigskip
\bigskip

\references

Arellano Ferro, A., Bramich, D.M., Figuera Jaimes, R., et
al. 2013, {\it MNRAS}, {\bf 434}, 1220

Bailey, S. I., 1902, {\it Harv. Ann.}, 38

Bailey, S. I., 1917, {\it Harv. Ann.}, {\bf 78}, 99

Bramich, D.M., 2008, {\it MNRAS}, {\bf 386}, L77

Bramich, D. M., Figuera Jaimes R., Giridhar S., Arellano Ferro A., 2011, {\it MNRAS},
{\bf 413}, 1275

Bramich D. M., Freudling W., 2012, {\it MNRAS}, {\bf 424}, 1584

Bramich, D.M., Horne, K., Albrow, M.D., Tsapras, Y., Snodgrass, C., Street, R.A., 
Hundertmark, M., Kains, N., Arellano Ferro, A., Figuera Jaimes, R. \& Giridhar, S.,
2013, {\it MNRAS}, {\bf 428}, 2275

Caputo, F., Castellani, V., Marconi, M., Ripepi, V.  1999, {\it MNRAS}, {\bf 306},
815

Clement, C.M., Muzzin, A., Dufton, Q., Ponnampalam, T.,
Wang, J., Burford, J., Richardson, A., Rosebery, T.,
Rowe, J., Sawyer-Hogg, H., 2001, {\it AJ}, {\bf 122}, 2587

Drissen, L., Shara, M. M. 1998, {\it AJ}, {\bf 115}, 725 

Gerashchenko, A. 1987, {\it IBVS}, 3044

Kadla, Z. I., Gerashchenko, A. N., Yablokova, N. V., Irkaev, B. N. 1987, 
{\it Ast. Tsirk.}, {\bf 1502}, 7

Kaluzny, J., Thompson, I., Krseminski, W., Pych, W. 1999, {\it A\&A}, {\bf 350}, 469

Oosterhoff, P. Th. 1941, {\it Leiden Ann}, {\bf 17}, Part 4

Stetson, P.B., 2000, {\it PASP}, {\bf 112}, 925
\endreferences

\end{document}